\documentstyle[aps,multicol,psfig]{revtex}

\newcommand{\tr}{{\rm tr}}
\begin{document}   
\bibliographystyle{prsty}
\widetext
\title{Non-universality of chaotic classical dynamics:
implications for quantum chaos}
\author{M. Wilkinson$^1$ and B. Mehlig$^2$
}
\address{
\mbox{}$^1$Department of Physics and Applied Physics,
University of Strathclyde,
Glasgow, G4 0NG, Scotland, U.K.\\
\mbox{}$^2$Theoretical Physics, 
University of Oxford, Oxford, OX1 3NP, England, U.K.}
\date{\today}
\maketitle{ } 
\begin{abstract}  
It might be anticipated that there is statistical universality in 
the long-time classical dynamics of chaotic systems, corresponding 
to the universal correspondence of their quantum spectral statistics 
with random matrix models. We argue that no such universality exists. We
discuss various statistical properties of long period orbits:
the distribution of the phase-space density of periodic orbits
of fixed length and
a correlation function of periodic-orbit actions, corresponding
to the universal quantum spectral two-point
correlation function. We show that bifurcations are a mechanism
for correlations of periodic-orbit actions. They lead
to a result which is non-universal, and which in general may not
be an analytic function of the action difference.
\end{abstract}   
\begin{multicols}{2}
It has been appreciated for many years that sufficiently
complex quantum systems exhibit a 
high degree of universality \cite{boh94}:
many statistical properties of their spectra usually fall
into one of three classes, exemplified by the three Gaussian
random matrix ensembles introduced by Dyson \cite{dys62}. It is natural
to ask whether an analogous degree of universality exists
in classical dynamics, and if it exists whether it underlies
the universality observed in the behaviour of quantum systems.
This paper suggests what the appropriate classical analog of
quantum spectral universality should be, and gives arguments
supporting the view that there is, in general, no classical
universality underlying that of quantum systems.

The universality exhibited by spectral properties is confined
to statistics which are sufficiently short ranged in energy.
This implies that the universal features are associated
with dynamics over long time scales: they may be associated
with universal properties of the long-time classical dynamics,
or they may be purely quantum.  
It is natural to anticipate that universal behaviour,
if present at all, will only be manifest in properties
which characterise small regions of phase space, since the
large scale structure of phase space can clearly be
non-universal.

\begin{figure}
\vspace*{-.4cm}
\centerline{\psfig{file=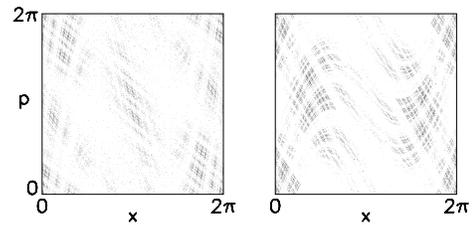,width=7cm}}
\caption{\label{fig:1} Shows  periodic 
points of the standard map with $K=6$ and $N=7$ (left)
and for the modified cat map $K=2$ and $N=11$ (right).
In both cases, $0 \leq x,p \leq 2\pi$.}
\end{figure}

An example of a property which could show universality
is the statistical characterisation of the distribution
of points periodic
under the $N^{\rm th}$ iterate of a chaotic 
area preserving maps of the form
$\alpha_{n+1} = {\cal M}(\alpha_n)$, with
$\alpha=(x,p)$.
These points are illustrated in Fig. \ref{fig:1} 
for the case of the $7^{\rm th}$ iterate
of the standard map \cite{chi79}
\begin{eqnarray}
\label{eq:1}
x_{n+1}&=&x_n+p_n\,,\\
p_{n+1}&=&p_n+K \mbox{sin}(x_{n+1})\nonumber
\end{eqnarray}
with $K=6$ and the $11^{\rm th}$ iterate
of a modified cat map
\begin{equation}
\label{eq:2}
\left(\begin{array}{l}
x_{n+1}\\
p_{n+1}\\
\end{array}\right)
=
\left(\begin{array}{ll}
1 & 1\\
1 & 2\\
\end{array}\right)
\left(\begin{array}{l}
x_{n}\\
p_{n}\\
\end{array}\right)
+
K
\left(\begin{array}{l}
u_n\\
0\\
\end{array}\right)
\end{equation}
with $u_n = \sin(x_n+2p_n)$ and $K=2$.
The patterns displayed in Fig.~\ref{fig:1}
are complex and show very wide fluctuations
in the density of points.
They are clearly different, but 
local statistical properties of their fine-scale structure might be
equivalent after scaling the coordinates to give the same mean
density of points. 

Gutzwiller's trace formula \cite{gutz90} 
gives a relation between classical periodic orbits and
quantum spectra. The formula is exact for a small number
of special systems, but it cannot be exact in general 
because it contains no information about the choice
of quantisation procedure \cite{wilk88}.
In \cite{arg93,smil97} Gutzwiller's formula is combined
with the observation that spectral
correlations are described by random matrix theory to infer 
that the actions of long period orbits are correlated. 
The correlations are a function of the action 
difference which was predicted to be universal within each of 
Dyson's symmetry classes, and an analytic form was quoted for
the GUE ensemble. In systems where Gutzwiller's
trace formula is exact (for a discussion of the
three known examples see \cite{gutz90,sch99})
such classcial correlations must certainly exist.
However, in general the trace formula is
not exact and it is thus necessary to find an entirely
classical mechanism for such correlations.
Up to now no such mechanism has been found.

The paper is organized as follows.
We will first argue that the long-time, local
structure of phase-space 
of chaotic systems is non-universal, using
theoretical arguments and numerical experiments.
We then characterize the remarkably
strong fluctuations of the density of
periodic points in Fig. \ref{fig:1} 
and provide theoretical arguments as well
as numerical evidence that these
fluctuations are log-normal.
Finally we propose an entirely classical
mechanism for periodic-orbit correlations,
based on the statistical properties
of bifurcations in the long-time dynamics.
This mechanism gives rise to a non-universal and possibly
non-analytic contribution to the correlation function.

We will first show that the long-time, local structure of phase space
is not universal. Our argument is based upon considering a
particular statistic: we will consider the proportion $P(N)$ of
trajectories which are elliptic upon $N$ iterations of
area-preserving maps such as (1), (2). The results generalise directly
to continuous time systems. 
The stability of a point periodic
under $N$ iterations of $\cal M$, $\alpha = {\cal M}^N(\alpha)$,
is described by the monodromy
matrix $M_N(\alpha) = \partial {\cal M}^N(\alpha)/\partial \alpha$ .
The point  $\alpha = {\cal M}(\alpha)$ 
is elliptic (hyperbolic) if
$|{\rm tr}\,M_N| <2$ ($|{\rm tr}\,M_N | >2$).
The terms elliptic and
hyperbolic will be used in the same way to describe any trajectory,
regardless of whether it is closed.

For chaotic maps, in the large-$N$ limit the elliptic and 
hyperbolic trajectories are
finely intermingled, implying that $P(N)$ is independent of any
smooth scaling of the phase-space coordinates, and $P(N)$
will decrease rapidly with increasing $N$.
Consider the distribution of values of ${\rm tr}\, M_N(\alpha)$
for large values of $N$. The monodromy
matrix $M_N$ is a product of elementary monodromy matrices
$m(\alpha) = M_1(\alpha)$
for individual applications of the mapping
$M_N(\alpha)=\prod_{n=1}^N m(\alpha_n)$
where $\alpha_n$ is the phase-space point reached from $\alpha$
after $n$ applications of the mapping. The typical
values of the elements of the matrix $M_N$ are expected to
grow exponentially, in the sense that the mean of the logarithm
of some norm of the matrix $M_N$ should grow linearly with
time: 
$\lambda={1\over 2}\lim_{N\to \infty} N^{-1}
\langle \log \tr\,[ M_N^T(\alpha)M_N(\alpha)] \rangle_\alpha$
is termed the Lyapunov exponent.
This suggests that elliptic trajectories are rare
in the long-time limit, and that if a universal form exists
for the function $P(N)$, it might be expected to be exponential,
\begin{equation}
\label{eq:3}
P(N)\sim A\exp(-a\,\lambda\, N)
\end{equation}
for $ N\gg 1$, where
$A$ and $a$ are universal constants.

Determing the fraction of elliptic trajectories is equivalent
to determining the probability that ${\rm tr}\, M_N$ lies
in the interval $[-2,2]$. Because the typical value of
${\rm tr}\, M_N$ is exponentially large, this question
relates to the tail of the distribution. In the case
of a chaotic map the succesive positions $\alpha_n$ have
the characteristics of random numbers, and the monodromy
matrix $M_N$ may be modelled as a product of random
matrices. A product of a large number of random scalars has
a log-normal distribution, and it is therefore natural
to expect that ${\rm tr}\, M_N$ will have a log-normal
distribution. The central limit theorem is only applicable
sufficiently close to the maximum of the probability distribution,
and the tails of the distribution of a sum of random
variables depends upon the distribution of the variables.
Because the distribution of the matrices $m(\alpha)$
is non-universal, we therefore expect that the tail of the
distribution of ${\rm tr}\, M_N$ is non-universal.
     
Further support for this prediction comes from considering
moments of $M_N(\alpha)$. Modelling the $\alpha_n$
as random variables, we have
\begin{equation}
\label{eq:4}
\langle M_N(\alpha)\rangle
=\prod_{n=1}^N \langle m(\alpha_n) \rangle
=\bigl[\langle m(\alpha) \rangle\bigr]^N
\end{equation}
where the averages are taken over the invariant measure
of the dynamics, which in the Hamiltonian case is just
the uniform distribution on the phase-space area. Higher
moments are obtained by averaging outer products
of monodromy matrices: 
$\langle \tr\, M_N^T(\alpha) M_N(\alpha)\rangle$
for example may be expressed 
in terms of 
$\bigl[\langle m(\alpha) \otimes m(\alpha)\rangle \bigr]^N$
(see also \cite{abr80}).
The value of $\langle \tr\, M_N^T(\alpha) M_N(\alpha)\rangle$
is therefore expected to grow exponentially with $N$
\begin{equation}
\label{eq:5}
\langle \tr\, M_N^T(\alpha) M_N(\alpha)\rangle \sim \exp(\lambda_2 N)
\end{equation}
where $\lambda_2$ is the largest eigenvalue
of $\langle m(\alpha) \otimes m(\alpha)\rangle$.
A similar approach
can be used to estimate the growth of
$\langle ({\rm tr}\, M_N^T M_N)^k\rangle$: the result
is of the form (5) with $\lambda_2$ replaced by $\lambda_{2k}$,
the largest eigenvalue of the $k$-fold outer product
$\langle m \otimes m \otimes \cdots \otimes m\rangle$. 
The values
of these eigenvalues depend upon the structure of the
elementary monodromy matrices $m$, indicating
that the ratios of the eigenvalues $\lambda_k$ are
non-universal. It is very difficult to reconcile this
with the hypothesis that $P(N)$ is universal.

\begin{figure}
\vspace*{-.8cm}
\centerline{\psfig{file=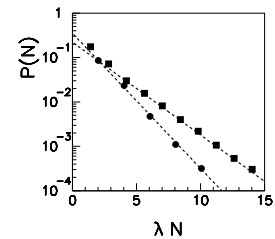,width=4.5cm}}
\caption{\label{fig:2} Shows $P(N)$ as a function
of $\lambda N$ for the standard map ($K=15$, $\lambda \simeq \log K/2 = 2.015$,
circles) and for the modified cat map ($K=8$, $\lambda = 1.404$, squares). 
}
\end{figure}

To test these predictions, the fraction of elliptic trajectories
and the Lyapunov exponent were evaluated for the mappings
defined by Eqs. (1) and (2).
In Fig.~\ref{fig:2} we plot the fraction $P(N)$ of elliptic trajectories
as a function of $\lambda N$. 
Fig.~\ref{fig:2} clearly shows
the non-universality of $P(N)$.

Next we discuss the
nature of the strong fluctuations apparent
in Fig.~\ref{fig:1}. This figure
indicates that  the density of periodic points is
highly non-uniform, however an exact result due to Hannay and
Ozorio de Almeida \cite{han84} suggests that the periodic points
might have a uniform distribution in phase space.
In the following, we discuss how this apparent contradiction
is resolved, and present an argument indicating that
the density of periodic points has a log-normal distribution.

We base our discussion on the Kac-Rice
\cite{kac32,rice44} approach to
calculating the densities of point singularities
of random functions. The problem is to find the
set of points where $\alpha={\cal M}^N(\alpha)$
for large $N$. The basis of the Kac-Rice approach is
to estimate the probability $\delta P$ of finding an
example of the singularity in a small ball of volume $\delta V$
centred on an arbitrary test point: this can be written
$\delta P\sim {\cal D}\,\delta V$, where ${\cal D}$ is the density of
singularities. The estimate of $\delta P$ is usually
simplified by the fact that the distance to a nearby
singularity is given by a simple expression.

Assume that there is a periodic point at a small
displacement $\delta \alpha $ from the test point
$\alpha $. Considering the condition for periodicity
$\alpha+\delta \alpha ={\cal M}(\alpha +\delta \alpha)$,
and expanding the mapping to first order in $\delta \alpha$
leads to the approximation
\begin{equation}
\label{eq:6}
\delta \alpha =(M-I)^{-1}(\alpha-{\cal M})
\end{equation}
(where the index $N$ designating the $N$-th
iterate has been dropped).
For long orbits, it may be assumed
that the joint distribution
function $P[{\cal M},M]$ factorizes, and
\begin{eqnarray}
\label{eq:7}
{\cal D}&=&P[{\cal M}]\,\,
\mbox{\rule[-3.5mm]{0.25pt}{7mm}\raisebox{-3mm}{\hspace{3pt}$
\scriptstyle{\cal M}=\alpha$}}\,
\int \!dM\,P[M]\,|\mbox{det}(M-I)|\\
&=& {\cal A}^{-1}\,\langle |\mbox{det}(M-I)|\rangle
\nonumber
\end{eqnarray}
where ${\cal A}$ is the phase-space area.
Similarly, the density of periodic points
weighted with a smooth function $W(\alpha)$ is
\begin{equation}
\label{eq:8}
{\cal D}_W={\cal A}^{-1}\,\langle W(\alpha)\,\vert {\rm det}(M-I)\vert
\rangle\,.
\end{equation}
Hannay and Ozorio de Almeida \cite{han84} considered the case 
$W(\alpha)=\vert w_j\vert^2$, where $w_j$ is
the weight of the $j^{\rm th}$
periodic orbit in the Gutzwiller sum.
This case is of importance when estimating quantum two-point
correlation functions semiclassically. 
The periodic orbit weights are of the form 
$w_j = \exp({\rm i}\pi \mu_j/2)\,\det(M-I)^{-1/2}$
where $\mu_j$ are integers termed Maslov indices
which account for the phase changes associated 
with focusing or reflection. This gives
$W(\alpha )=\vert {\rm det}( M- I)\vert^{-1}$, so
that in this case ${\cal D}_W={\cal A}^{-1}$,
which constitutes the version of the sum rule derived in
\cite{han84} which is applicable to maps. This simple, universal
form is a result of the cancellation of the weight
$\vert {\rm det}(M-I)\vert$ against its inverse.

\begin{figure}
\vspace*{-1.cm}
\centerline{\psfig{file=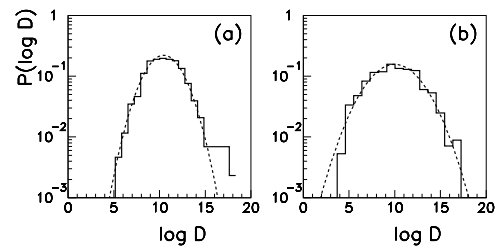,width=7cm}}
\caption{\label{fig:3} Shows the distribution
of the local density of periodic orbits ${\cal D}(\alpha)$ for
the standard map with $K=6$ and $N=7$ (a)
and for the modified cat map with $K=2$ and $N=11$ (b).
In both cases the distribution is log-normal
$(- - -$).}
\end{figure}

We can now comment on the highly non-uniform distribution of
periodic points shown in Fig.~\ref{fig:1}. It is natural to
attempt to characterise this by a local density
${\cal D}(\alpha )$ defined in terms of the number of orbits
inside a ball of radius, say,  $\epsilon $.
In order to construct a
satisfactory definition using this approach, the local
density would have to converge over a range of values
of $\epsilon$, this range becoming broader as $N$
increases. The fluctuations in density are so wild that
it appears to be impossible to define a local density
in this way. Instead, we {\sl define} the local density
\begin{equation}
\label{eq:9}
{\cal D}(\alpha)=\vert \delta \alpha \vert^{-d}\,.
\end{equation}
where $\delta \alpha$ is the displacement from the point
$\alpha $ to the nearest periodic point, $\vert \delta \alpha \vert$
is the corresponding Euclidean distance, and $d$ is 
the phase-space dimension. For points sufficiently
close to a periodic point, we have already seen that
$\vert \delta \alpha \vert \sim \vert (M-I)^{-1}
(\alpha -{\cal M}(\alpha))\vert$. The fluctuations of this quantity
are dominated by those of $M$ which are log-normal.
We plot the distribution of the local density evaluated
according to (\ref{eq:9}) in Fig.~\ref{fig:3}. 
It is well described by a log-normal distribution.

We now discuss the implications of our findings
for a minor variation upon a classical correlation function
of periodic-orbit actions first
introduced in \cite{arg93}
\begin{eqnarray}
\label{eq:10}
C(T,\Delta S)&=&\sum_{j\ne j'} w_j\, w_{j'}^\ast\,
\delta_{\eta} (T-T_j)\\
&\times&
\delta_\epsilon \big(\Delta S-(S_j-S_{j'})\big)
\,.
\nonumber
\end{eqnarray}
Here, $S_j$ and $T_j$
are the action and period of
the $j^{\rm th}$ periodic orbit. 
In systems where Gutzwiller's trace formula
is exact, and which exhibit universal
spectral two-point correlations,
$C(T,\Delta S)$ must be a 
function of the action difference which
is universal within each of Dyson's symmetry classes \cite{arg93,smil97}.
In general, however, the trace formula cannot be exact.
It is thus desirable
to explore classical models
for periodic-orbit correlations contained in (\ref{eq:10}).

A mechanism for such correlations
is revealed by embedding the Hamiltonian (or mapping) into a 
one-parameter family. We remark that systems for
which exact trace formulae exist (Riemann's zeroes,
motion of a surface with constant negative
curvature \cite{gutz90} and quantum graphs \cite{sch99})
do not form one-parameter families exhibiting
bifurcations.
The parameter $X$ could be a coupling
constant such as $K$ in Eq. (\ref{eq:1}).
Varying the  parameter $X$ produces bifurcations of orbits: 
immediately after
a bifurcation the two orbits have the same actions $S_j=S_{j'}$,
and their weights $w_j$ and $w_{j'}$ may be related
in a simple way (typically they have opposite signs).
We will estimate the contribution from bifurcations
to the correlation function (\ref{eq:10})
at small values of $\Delta S$,
as this can be directly
related to the forms of the bifurcations. Assume that a bifurcation
occurs at $X_i$ as a parameter $X$ is varied.
The action difference
in the neighbourhood of the bifurcation is of the form
\begin{equation}
\label{eq:11}
\vert \Delta S_i\vert \sim v_i \vert \Delta X_i\vert^\beta \,,
\end{equation}
where $\Delta X_i=X-X_i$.
The exponent $\beta$ will be determined from the type of bifurcation,
and the constant $v_i$ depends on the particular
bifurcation. 
The weights $w_j$ in (\ref{eq:10}) may
be either singular or regular in the neighbourhood
of the bifurcation: we allow for them having an algebraic
singularity with exponents $\gamma_j$, which may be different
for the two orbits in the vicinity of
the bifurcation,
\begin{equation}
\label{eq:12}
w_j\propto \vert \Delta X_i\vert^{\gamma_j}\,.
\end{equation}
In the following we discuss a model for evaluating
the contributions to (\ref{eq:10}) for long periodic orbits,
treating the $w_j$ in (\ref{eq:10}) as random variables,
which are independent except
for those pairs of orbits which are related by a bifurcation:
\begin{equation}
\label{eq:13}
\langle w_j \rangle =0\ ,\ \ \
\langle w_j w_{j'}\rangle = w^2\,\delta_{jj'}
\end{equation}
where in the latter case it is assumed that the orbits $j$ and
$j'$ are not related by a bifurcation.
At a bifurcation,
one or both of the weights may be singular.
In this case, 
it is reasonable to model the behaviour of $\langle w_j w_{j'}\rangle$
as follows
\begin{equation}
\label{eq:14}
\langle w_jw_{j'}\rangle \sim w^2\,
\Lambda^{-(\gamma_j+\gamma_{j'})}
\vert \Delta X\vert^{\gamma_j+\gamma_{j'}}
\end{equation}
where $\Lambda$ characterizes the frequency
with which bifurcations occur.

Let $\delta P_\epsilon(\Delta S)$ be the probability that a 
periodic orbit is connected to another orbit for which the
action difference is in the interval $[\Delta S,\Delta S+\epsilon]$.
According to the model above, the only contributions to the 
correlation function come from pairs of periodic orbits related 
by a bifurcation. The contribution from these orbits may be
estimated as follows
\begin{equation}
\label{eq:15}
C\sim -{1\over{\epsilon}}\biggl({dN\over{dT}}\biggr)
w^2\biggl({\Delta X\over{\Lambda}}\biggr)^{(\gamma_j+\gamma_{j'})}
\delta P_\epsilon (\Delta S)
\end{equation}
where $N(T)$ is the number of periodic orbits with period
less than $T$, and $\Delta X$ is the typical distance to a 
bifurcation. From (\ref{eq:11}), the bifurcation occurs 
at a displacement in parameter space
of the form $\Delta X \sim (\Delta S/v)^{1/\beta}$.

We model the probability that a periodic orbit does not
undergo a bifurcation in a distance $\Delta X$ from an
arbitrarily chosen test point as a Poisson distribution:
this probability is 
$P_\pm (\Delta X)=\exp [-\Lambda_\pm \vert \Delta X\vert]$
for displacements to either side of the test point.
For a sufficiently small separation $\Delta X$, the
probability of finding a bifurcation in a small interval
of size $\delta \Delta X$ on either side of $X_0$ is
\begin{equation}
\label{eq:16}
\delta P\sim \Lambda\,\delta \Delta X
\end{equation}
with  $\Lambda = \Lambda_++\Lambda_-$.
Combining (\ref{eq:11}) and (\ref{eq:16}),
one obtains $\delta P_\epsilon (\Delta S)\sim \Lambda v^{1/\beta}
\vert \Delta S\vert ^{1/\beta-1}\delta \Delta S$. Identifying 
$\delta \Delta S$ with $\epsilon $, and substituting into 
(\ref{eq:15}) produces the following result, valid for small 
$\vert \Delta S\vert$:
\begin{equation}
\label{eq:17}
C(T,\Delta S) \propto \vert \Delta S\vert
^{(1+\gamma_j+\gamma_{j^\prime}-\beta)/\beta}\,.
\end{equation}
Evaluation of the exponent in (\ref{eq:17}) requires 
information about the nature of the bifurcations.
As an example, consider a hyperbolic billiard system
with no corners, such as the Lorentz gas. In this case 
inverse bifurcations occur when pairs of similar orbits, only one 
of which bounces off a surface, become tangential
to that scattering surface. Geometrical considerations imply that 
$\beta=2$, and $\gamma_j=0$ for the orbit which does not bounce
tangentially but $\gamma_{j'}=1$ for that which does. In this
case the exponent in (\ref{eq:17}) is zero, but in other cases
the exponent may be non-zero.

We summarise the conclusions on correlations of periodic orbits 
as follows. In systems with a smooth 
Hamiltonian, the statistic $P(N)$ is very closely related to the 
condition for bifurcations (namely that a periodic orbit intersects
the manifold where $\tr M=\pm 2$). The statistic $P(N)$ was shown 
to be non-universal, and it is implausible that bifurcations will 
show universal behaviour in such systems. 
Totally chaotic systems have no elliptic trajectories, and bifurcations 
are associated with singularities of the Hamiltonian: in this case
the arguments against universality are no less compelling. 
We therefore conclude that there
is a non-universal component to the periodic orbit correlation
function. Our result (\ref{eq:17}) also indicates that the 
contributions to the correlation function coming from bifurcations
has a singularity if the exponent $1+\gamma_j+\gamma_{j'}-\beta$
is non-zero. It is difficult to conceive of a non-universal and
non-analtyic contribution to the correlation function which
could combine with this one to give a universal result.

\narrowtext
\end{multicols}
\end{document}